\pdfoutput=1
\documentclass{JINST}
\let\ifpdf\undefined
\usepackage{textcomp} 
\DeclareGraphicsExtensions{.png,.pdf} 
\usepackage{ifpdf}

\title{The ALICE EMCal L1 trigger first year of operation experience}

\author{O. Bourrion$^a$\thanks{Corresponding author.},
N.~Arbor$^a$,
G.~Conesa-Balbastre$^a$,
C.~Furget$^a$,
R.~Guernane$^a$
and G.~Marcotte$^a$ on behalf of the ALICE-EMCal collaboration\\
\llap{$^a$}Laboratoire de Physique Subatomique et de Cosmologie,\\ 
Universit\'e Joseph Fourier Grenoble 1,\\
  CNRS/IN2P3, Institut Polytechnique de Grenoble,\\
  53, rue des Martyrs, Grenoble, France\\
  E-mail: \email{olivier.bourrion@lpsc.in2p3.fr}}

\abstract{The ALICE experiment at the LHC is equipped with an electromagnetic calorimeter (EMCal) designed to enhance its capabilities for jet, photon and electron measurement.
In addition, the EMCal enables triggering on jets and photons with a centrality dependent energy threshold. 
After its commissioning in 2010, the EMCal Level 1 (L1) trigger was officially approved for physics data taking in 2011. 
After describing the L1 hardware and trigger algorithms, the commissioning and the first year of running experience, both in proton and heavy ion beams, are reviewed. 
Additionally, the upgrades to the original L1 trigger design are detailed.}


\begin{document}

\section{Overview}
ALICE (A Large Ion Collider Experiment)\cite{ALICEref} at the LHC\cite{LHCref} is a general purpose experiment designed to study the phase transition between ordinary nuclear matter and the quark-gluon plasma, which occurs in high energy nucleus-nucleus collisions. 
To enhance its capabilities for measuring jet properties, the ALICE detector has been upgraded in 2010 with a large acceptance ($\Delta \eta \times \Delta \phi = 1.4 \times 1.86$ (107\textdegree)) ElectroMagnetic Calorimeter (EMCal)\cite{EMCALTDR} providing a measurement of the neutral fraction of the jet energy and an unbiased jet trigger, thanks to a centrality dependent energy threshold. 

The sampling calorimeter consists of 12288 towers of layered Pb-scintillator arranged in modules of $2 \times 2$ towers, with each tower containing 77 layers for a total of 20.1 radiation lengths.
A tower is read out with an avalanche photodiode (APD) which collects, via a bundle of optical fibers, the light created by particle interactions.
A charge sensitive preamplifier (CSP) is used to instrument each APD.
A supermodule (SM) is made of 24 strips of 12 modules (1152 towers). In 2010, the EMCal comprised four SM, in 2011 there were ten SM and in 2012 the full EMCal contains ten complete SM plus two thirds of a SM.


\begin{figure}
  \begin{center}
   \includegraphics[angle=0,width=0.5\textwidth]{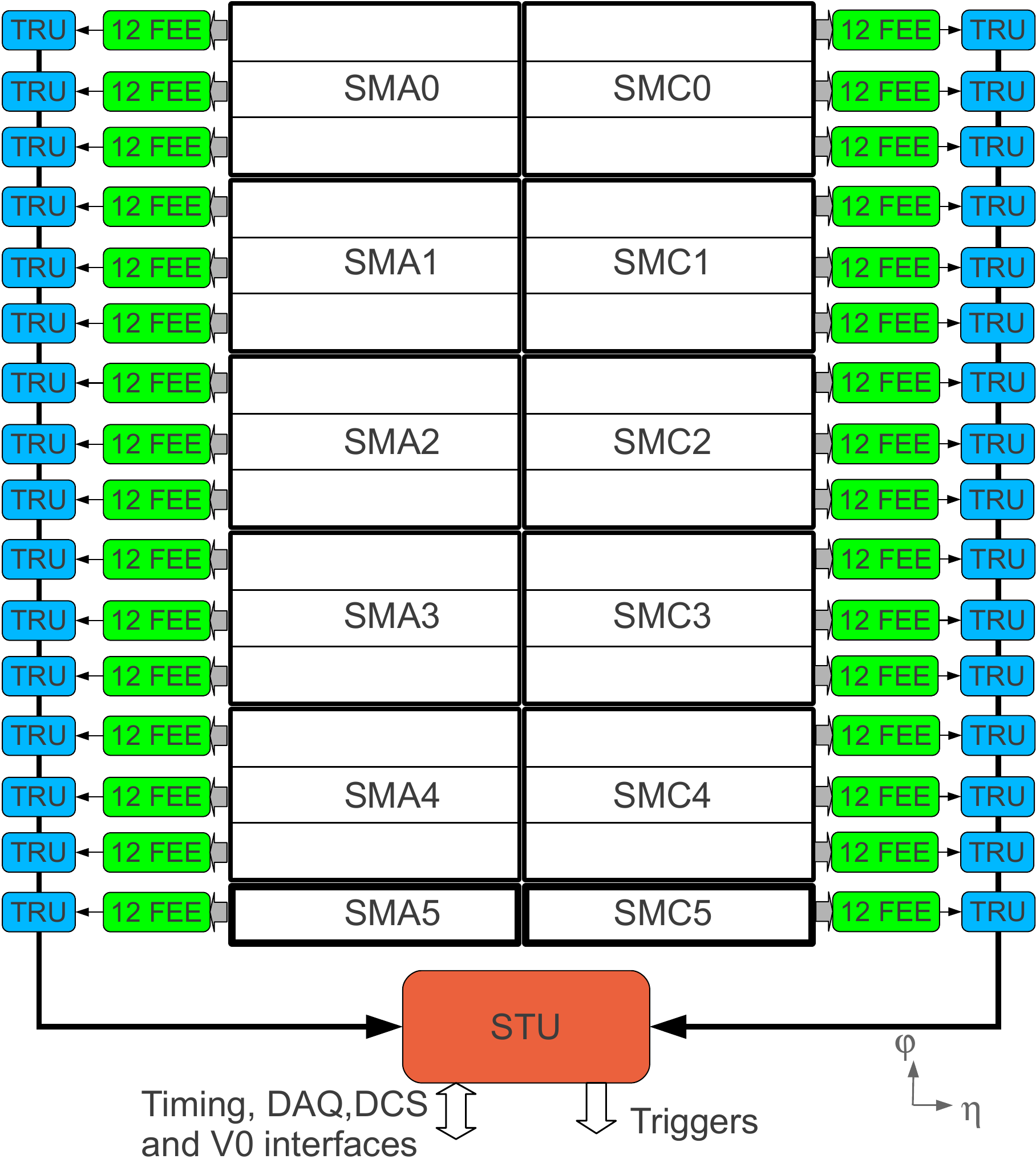}
  \caption{Flat view of the EMCal detector with its surrounding trigger electronics.}
  \label{elec_overview}
  \end{center}
\end{figure}
A schematic view of the EMCal detector with its front-end and trigger electronics is sketched in fig.~\ref{elec_overview}. 
Each SM is divided into three regions, and each region is instrumented by 12 FEE cards, so each SM has 36 FEE cards\cite{FEEpaper}.
Each FEE takes 32 analog inputs and generates eight fastOR signals. These are fast shaped (100\,ns) analog sums over one module, i.e. four tower signals.
The individual tower signals are used for energy measurement while the 3072 module analog sums (fastOR), are used to build the trigger. 
The Trigger Region Unit (TRU) \cite{TRUpaper} are used to digitize, at the machine bunch crossing rate (40.08\,MHz), the fastOR signals provided by the FEE and to compute and generate the local Level 0 (L0) triggers. Finally, the Summary Trigger Unit (STU), computes the global L0 trigger by ORing the local L0 triggers.
The STU also collects and aggregates the TRU data used to compute the the two  Level 1 (L1) triggers, the photon trigger and the jet trigger.
The L1 thresholds are computed event-by-event using the ALICE beam-beam counter detector\cite{V0paper} (V0) according to a 2\textsuperscript{nd} order fit function $A \cdot V0_{count}^2 + B \cdot V0_{count}+C$, where $V0_{count}$ is the total charge information provided by the V0 and A, B, C the threshold parameters.

The communication between the TRUs and the STU is performed through 12\,m point to point cat7 Ethernet cables.
Additionally, the STU is included in the EMCal readout via a Detector Data Link\cite{DDLpaper} (DDL) to the ALICE DAQ\cite{ALICEDAQ}. The readout, which is primarily used to return the triggering indexes and thresholds used on a event-by-event basis, can also be used to provide the primitive triggering data in order to recheck off-line the on-line trigger quality. 
Additionally, an Ethernet interface to the Detector Control System (DCS) interface is used for fast FPGA firmware upload and run configuration (thresholds parameters, trigger delays, etc).

\section{Trigger algorithms}
\subsection{TRU L0 algorithm}
After digitization, each fastOR is digitally integrated over a sliding time window of four samples. Then, the results of these operations are continuously fed to $2 \times 2$ spatial sum processors that compute the energy deposit in patches of $4\times4$ towers (or $2 \times 2$ fastOR) for the region managed. Each patch energy is constantly compared to a minimum bias threshold; whenever it is crossed and the maximum of the peak has been found, a local L0 trigger is fired. In preparation for the L1 algorithm, the time integrated sums are also stored in a circular buffer for later retrieval and transmission to STU.
Note that Level 0 trigger suffers from some spatial trigger inefficiencies due to the fact that TRUs cannot compute the spatial sum for patches sitting on region boundaries

\subsection{Global EMCal triggers computed in STU}
\begin{figure}[b]
  \begin{center}
   \includegraphics[angle=0,width=0.8\textwidth]{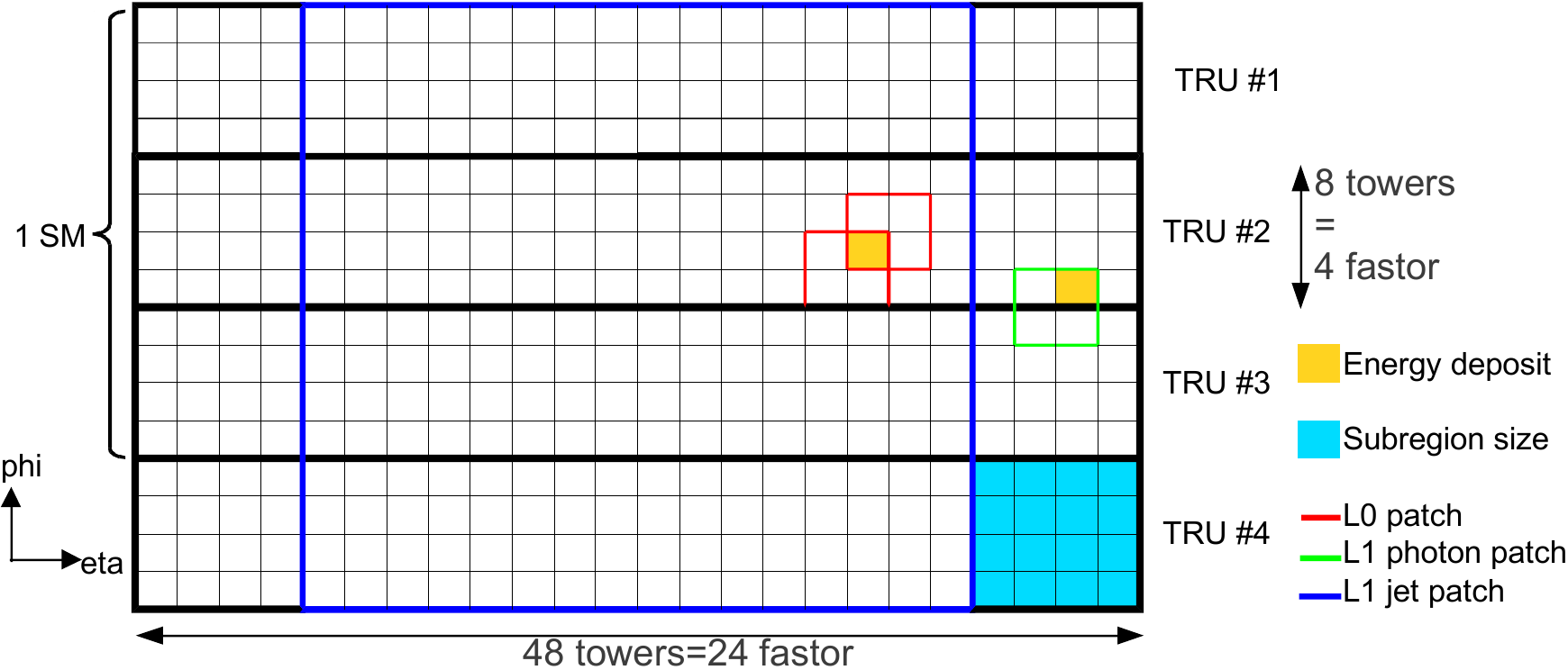}
  \caption{Cartoon of different possible L0, L1-gamma and L1-jet trigger patches.}
  \label{SM_map}
  \end{center}
\end{figure}

The STU is the access point to the Central trigger Processor CTP\cite{CTPpaper} for EMCal. 
Consequently, it is used to provide the global L0, which is an OR of the 32 L0 locally calculated by the TRUs and two L1 triggers: the L1-gamma trigger and the L1-jet trigger.
The L1-gamma trigger uses the same patch size as L0, but without the inefficiencies displayed by the local L0 (i.e. $2\times2$ patch across several TRU regions can be computed).
The L1-jet trigger is built by summing energy over a sliding window of $4\times4$ subregions, where a subregion is defined as a $4 \times 4$ fastOR (or $8 \times 8$ towers) area, see fig.~\ref{SM_map}.
With the given EMCal geometry and due to the various trigger patches sizes, there are a total of 2208 L0, 2961 L1-gamma and 117 L1-jet trigger patches that can be fired.

\subsection{L1 trigger processing}
\begin{figure}
  \begin{center}
  \includegraphics[angle=-90,width=0.9\textwidth]{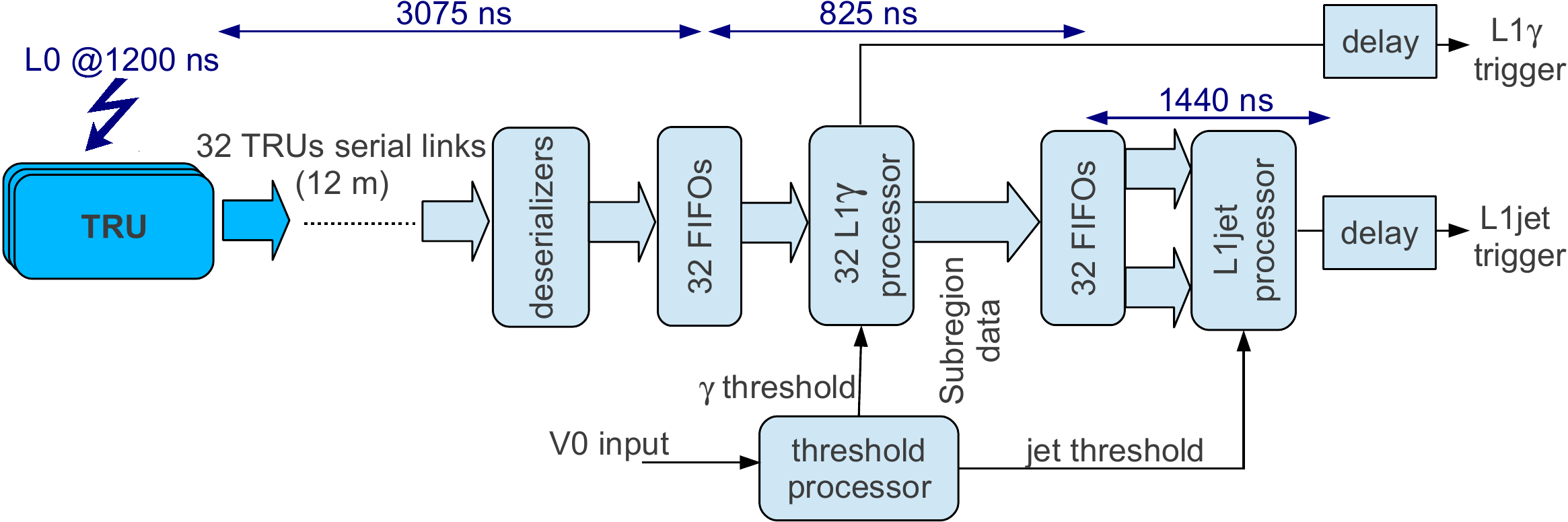}
  \caption{Block diagram of the L1 trigger processing annotated with the time required to go through each step. }
  \label{L1_trig_proc}
  \end{center}
\end{figure}

A block diagram of the L1 trigger processing is shown in fig.~\ref{L1_trig_proc}.
The L1-processing is not continuously running, i.e. pipelined, it is instead initiated on the confirmed L0 reception provided by the CTP via the TTC\cite{TTCpaper} links (TRUs and STU).
At this moment, 1.2\,\textmu s after interaction, the TRUs send to the STU the appropriate time integrated data from their circular buffers to the STU via the custom serial links. 
The serialization, propagation delay and deserialization takes 3075\,ns.
Meanwhile, the V0 detector transfers its charge information to the STU via a direct optical link. The thresholds for photon and jet patches are immediately processed and made available before the actual trigger processing starts.
Once the TRU data reception is achieved, the L1-photon trigger processing and also the subregion energy calculation are done in parallel for each TRU region.
Then when the previous processing is over, the L1-jet trigger starts and uses the previously generated subregion accumulation. Finally, both triggers are adequately delayed to accommodate the L1-trigger latency expected by the CTP.
More technical details about the trigger implementation may be found in \cite{STU_twepp2010}.

\section{Custom serial protocol}
\subsection{Original solution}
\begin{figure}
  \begin{center}
   \includegraphics[angle=-90,width=0.85\textwidth]{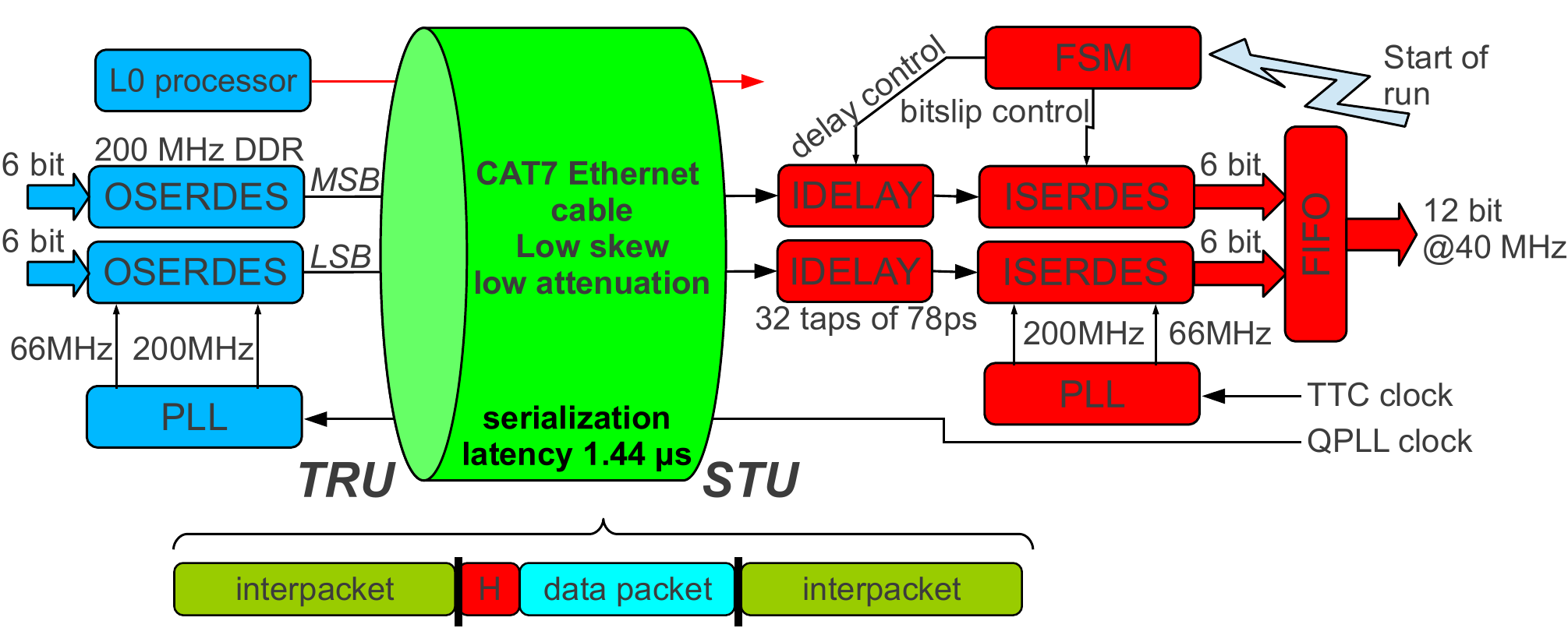}
  \caption{Sketch of the original custom serial link.}
  \label{original_serial_link}
  \end{center}
\end{figure}

The main motivation for the development of this custom serial link was the desire to reuse the TRU design made for the \textbf{PHO}ton \textbf{S}pectrometer (PHOS) which was equipped with a spare RJ45 connector directly linked to its FPGA.
The original trigger timing constraints drove the design in the same direction.
This solution minimizes transmission latency and meets some functional requirements, allowing the STU to be used as a low jitter reference clock distributor for TRUs.
Additionally, the fact that the local L0s had to be forwarded to STU for feeding its global OR required a custom solution.
Thus, the choice was made to use a four-pair LVDS link transported over CAT7 Ethernet cables because they have the appropriate impedance and feature low signal attenuation and low skew between pairs (see fig.~\ref{original_serial_link}).
Pair usage is as follows: one pair is dedicated for the LHC reference clock transfer to the TRU, another is used by the TRUs to forward their local L0 candidates and the two remaining are used for synchronous serial data transfer without any encoding.
Each data pair was running at 400\,Mb/s and the clock used for transfer is the LHC clock multiplied by 10.
With this very light protocol, the latency is only the sum of the cable delay and bit transmission time.
Each TRU sends simultaneously its 96 values of 12 bit coded time integrated fastOR data to the STU at 800\,Mb/s; in this case the serialization latency is thus 1.44\,\textmu s. The communication protocol was simple, outside of the data payload transmission a known inter-packet word was continuously transfered. Then at the transmission time, right after the confirmed L0 reception, a header packet was sent followed by the time-integrated data. 

The link synchronization is done before each start of run by a Finite State Machine (FSM) implemented in the FPGA. 
This is done in two steps. In the first step, the data phase alignment takes place, it relies on the individual data path fine granularity delaying feature available in the Virtex 5 FPGA (up to 64 steps of 78\,ps).
A scanning of all delay values is made in order to obtain the zone where data reception is stable and then the central value is applied.
In the second step, character framing is performed for associating individually each incoming bit to the good deserialized word.
This whole process is performed with the inter-packet data word used as the synchronization/training pattern.

For the link quality monitoring, error counter monitors were implemented.
These were incremented for each bad inter-packet received outside of the expected payload transmission.
The counters were checked every minute via DCS, and an alarm was raised in case of  transmission errors.

\subsection{Problem encountered and diagnosis tool}
The original custom serial link solution was successfully validated in the laboratory and also in 2010 with four installed SM by regularly performing TRU/STU data correlation checks. Unfortunately, in 2011, when the EMCal was fully installed, several random TRU-STU links were displaying communication errors during some runs, while for all links the start of run synchronization went through correctly.
As expected from the missing links, off-line validation showed missing L1-photon triggers for the missing regions. 
But, from time to time, the on-line L1 jet trigger rate (relative to accepted L0 triggers) jumped from a nominal value of 2\% to 100\% (no rejection). 

\begin{figure}
  \begin{center}
   \includegraphics[angle=-90,width=0.95\textwidth]{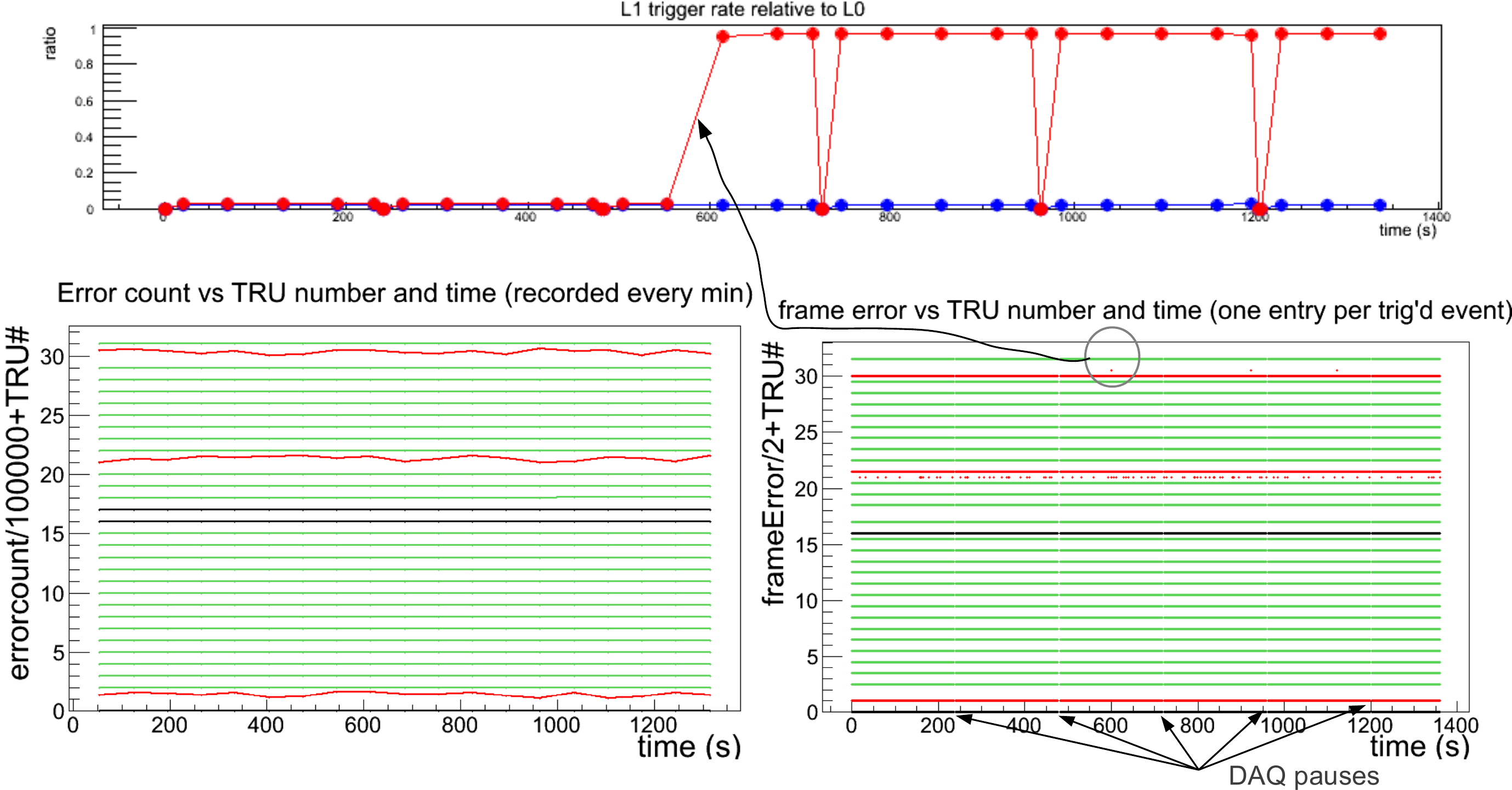}
  \caption{Communication failure and trigger rate diagnosis  of run 163532. 
  Top figure shown the L1 trigger rate relative to accepted L0 trigger (L1-jet in red and L1-photon in blue). This plot was obtained by dividing the error count (maximum of 65535) by 100000 and by adding the corresponding TRU number.
  Bottom left plot shows the error count recorded for every minute and for each TRU. Bottom right plot shows the frame bit received for each accepted trigger and for each TRU. This latter plot was obtained by dividinng the frame error (maximum of 1) by 2 and by adding the corresponding TRU number.
  The correlation between the first successful communication of TRU-STU link 30 and the L1-jet trigger rate increase is shown.}
  \label{frame_errors_vs_rate}
  \end{center}
\end{figure}

In order to understand where the problem lay, a frame reception monitor was inserted in the deployed firmware.
It is able to check, for each TRU-STU link and for each event, the good or bad reception of the packet header.
The resulting reception bit mask is inserted in the data stream along with the corresponding event.
A run diagnosis example is shown in fig.~\ref{frame_errors_vs_rate} for  run 163532. It can be seen that while the error counter monitoring tool is indicating that TRU 1, 21 and 30 are badly communicating with the STU, the frame reception monitor shows that in fact TRU~1 is not communicating at all with the STU and that TRU~21 is transmitting data most of the time. Remarkably, TRU~30 has only three successful data transfers toward STU and the first one is actually causing the trigger rate increase. 
This observation not only confirmed the suspected communication problem, but also revealed that there was a flaw in the L1-jet trigger algorithm implementation.
For fixing the communication problem, the first cure attempt was to decrease the transmission rate to $2 \times 240$\,Mb/s, thus relaxing the serial link timing constraints. This was possible in 2011, thanks to the increased timing budget for providing the candidate L1 trigger at the CTP input (from 6.2 to 7.3\,\textmu s).
Unfortunately, this did not solve the problem.
By performing a data recording at fixed latency after confirmed L0 reception, instead of recording the payload after a packet header reception, the issue was found to be due to the serialization/deserialization.
While the synchronization seemed good, sometimes a cycle delay between the LSB part and MSB part of the transmitted data word appeared. Obviously, this problem could not be observed at the synchronization time with a single word training pattern.
Therefore, the second, and successful, cure applied was to use a three word training pattern, in conjunction with the possibility to delay the MSB or the LSB part of a word during the synchronization phase.

\section{Correcting fake and missing triggers fixing, from simulation to on-line debugging}
\begin{figure}
  \begin{center}
   \includegraphics[angle=-90,width=0.6\textwidth]{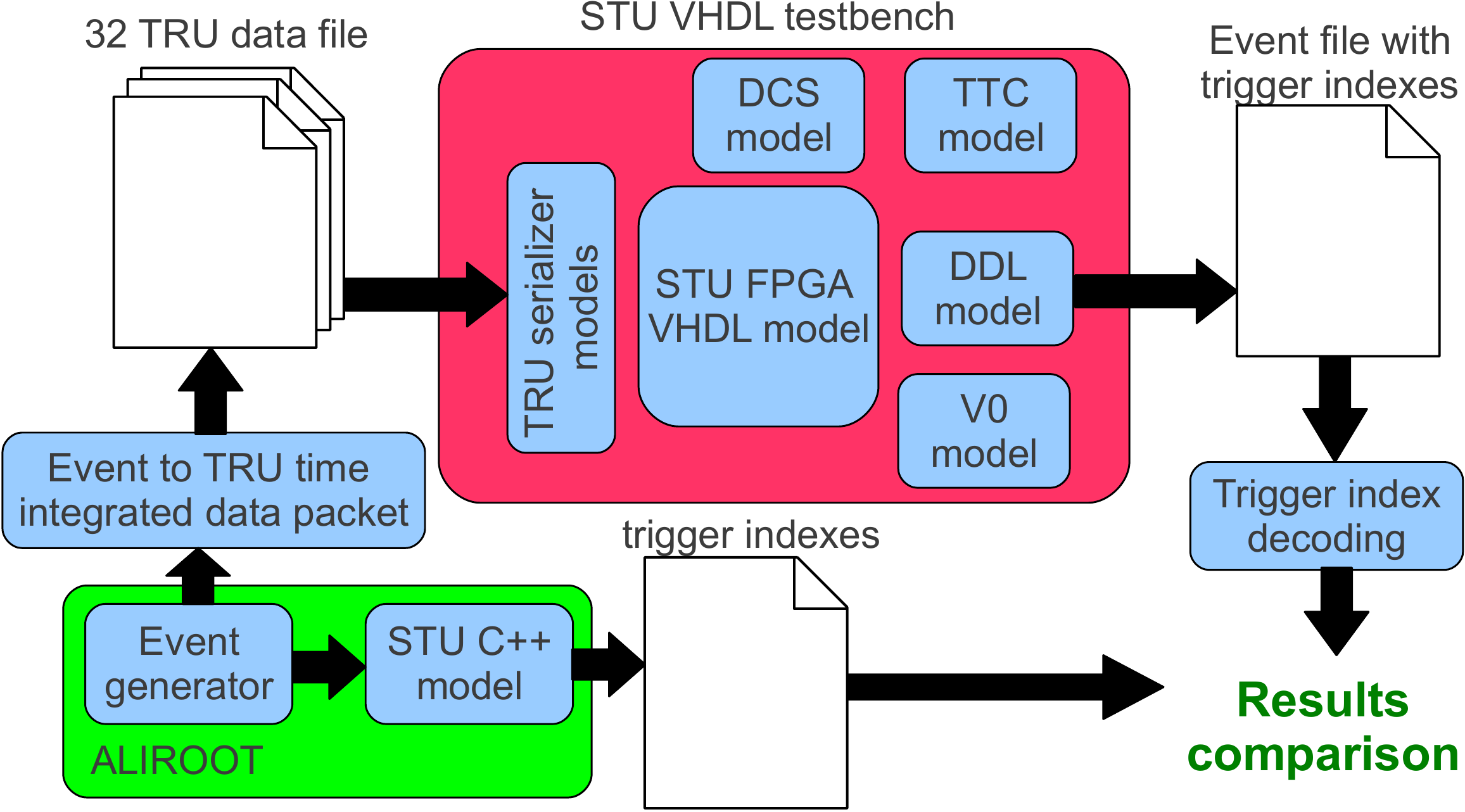}
  \caption{Overview of the ``physics'' and ``firmware'' co-simulation.}
  \label{vhdl_aliroot}
  \end{center}
\end{figure}

From the early development stage of the hardware and firmware, gateway tools were developed to exchange data between ``physics and ''firmware`` simulations, as shown schematically in fig.~\ref{vhdl_aliroot}.
This allowed for the validation of the core STU algorithms (jet and photon) before deployment and, as a side benefit, for the quick adaptation of gateway tools --- such as the trigger index decoding routine --- to the off-line software.

\begin{figure}
  \begin{center}
   \includegraphics[angle=-90,width=0.8\textwidth]{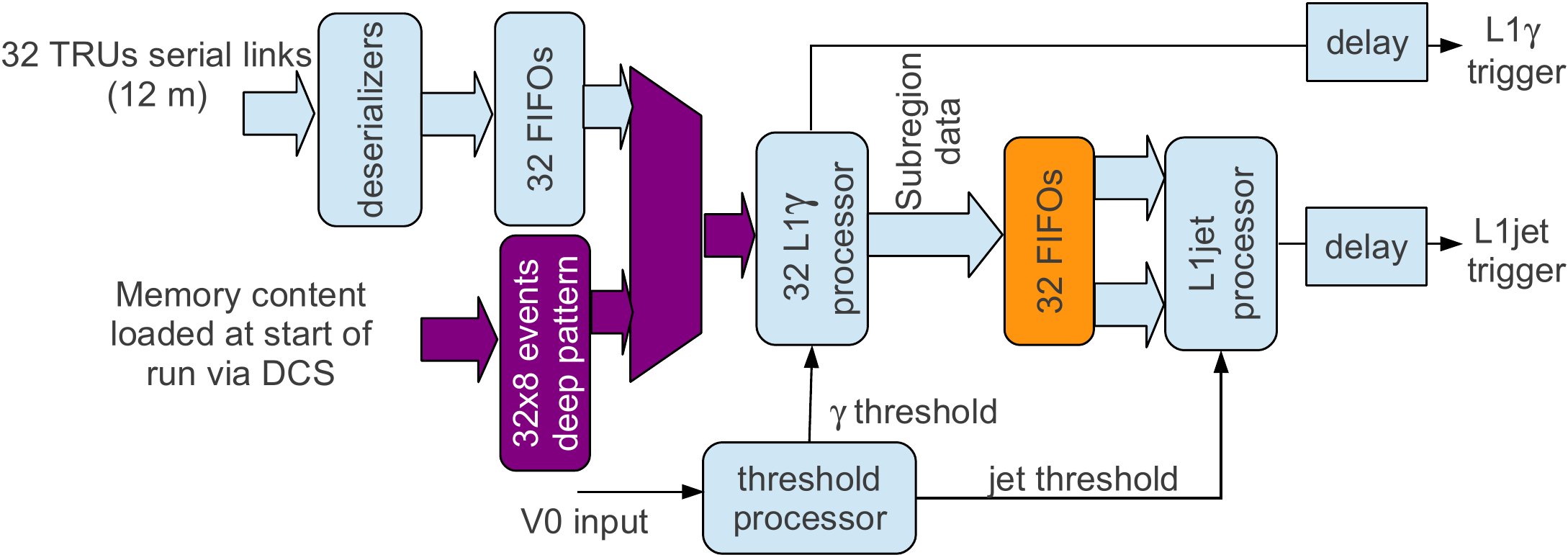}
  \caption{Modified STU firmware featuring the ``event-player'' allowing to select the data source to be used by the trigger processor between the TRU received data and DCS preloaded data. The buffer causing fake L1-jet trigger when not reseted between event and with flaky communication is colored in orange. }
  \label{L1_trig_proc_pattern}
  \end{center}
\end{figure}

While these tools were useful in the early stage of development, they were limited. 
For instance, the firmware simulation is slow. Moreover, it is not easy to validate all possible external effects which could cause false and/or missing triggers.
Examples of such possible effects include communication breakdown, clock jitter, and other, not necessarily predictable, issues.
Consequently, an ``event player'' feature was added in the STU firmware.
As shown in fig.~\ref{L1_trig_proc_pattern}, this on-line tool allows to select the data to be used by the trigger processors between the TRU received data and DCS preloaded data.
The ``event player'' can play up to eight different patterns. 
It offers the possibility of validating the entire L1 algorithms in-situ and to check the compliance with the ALICE DAQ after each data packet modification.  
Additionally, it may be used to accumulate statistics to check for eventual timing issues or other such as radiations effect.
Thanks to this debugging tool, the L1-jet trigger rate issue was pinpointed to the missing ``sub-region'' buffer clearing when serial communication links failed. 
Indeed, for flaky links, the ``sub-region'' computed from the last correctly received data were constantly used by the L1-jet processor.
Hence, when the last received information contained a high energy event, the subsequent L0 confirmed event were mistakenly accepted at L1.
After correcting this problem, both L1 triggers performed as expected, as detailed in the next section. 

\section{Trigger performance}
\begin{figure}
  \begin{center}
   \includegraphics[angle=0,width=0.4\textwidth]{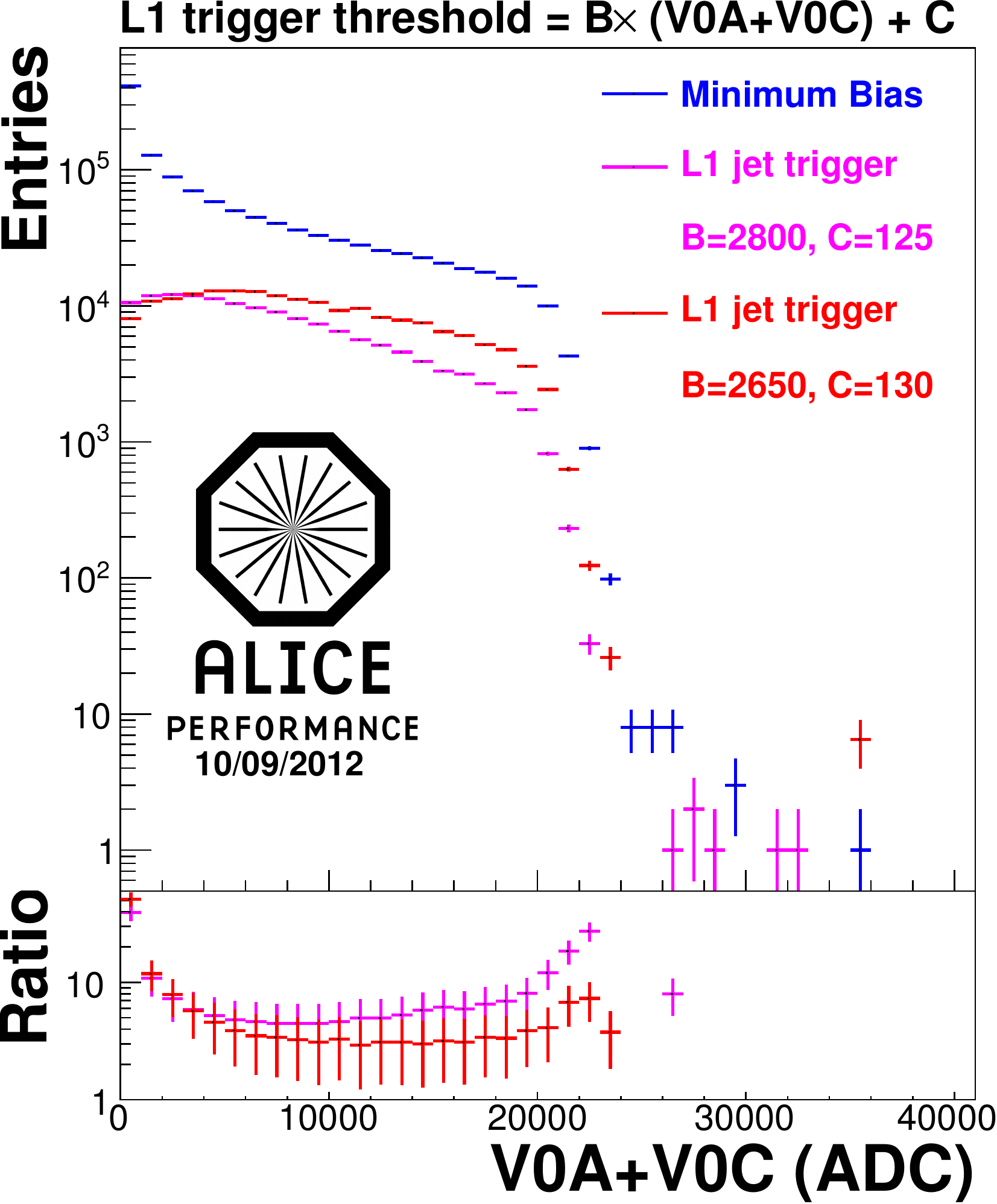}
  \caption{Plot of the event selection for Pb/Pb in 2011. The event yields versus the centrality for minimum bias and EMCal L1-jet trigger classes is shown. 
  The lower plot shows the MB to L1-jet ratios distribution for the different threshold parameters. Horizontal scale is the total amount of V0 charge expressed in ADC counts.}
  \label{plot_efficacite}
  \end{center}
\end{figure}

As an illustration of the trigger performance during the 2011 lead beam period (Pb-Pb collisions at $\sqrt{s_{NN}}$ = 2.76\,TeV), the event selection as a function of the centrality of the collision for Minimum Bias (MB) and EMCal L1-jet trigger classes is shown in fig.~\ref{plot_efficacite}. 
The upper plot shows the minimum bias\footnote{The minimum bias trigger class is composed of the coincidence of the V0 detector L0 trigger signal and the ZDC L1 trigger signal (Zero Degree Calorimeter).} and L1-jet samples\footnote{The L1-jet sample is a subsample of the MB, obtained with the coincidence with EMCal L1 jet trigger signal.} for a linear energy threshold with two threshold parameter sets.
The lower plot shows the MB to L1-jet ratios for the different threshold parameters. 
The set of parameters giving the magenta distribution rejects too many central events, while the set of of parameters giving the red one is more uniform for V0A + V0C signal above 5000 ADC.
The L1 trigger could provide a uniform background rejection, in a large centrality region, while disfavoring the most peripheral events.
This behavior, inherent to the order of the threshold computation, will be improved by using a second order centrality dependent energy threshold.

\begin{figure}
  \begin{center}
  \includegraphics[angle=0,width=0.48\textwidth]{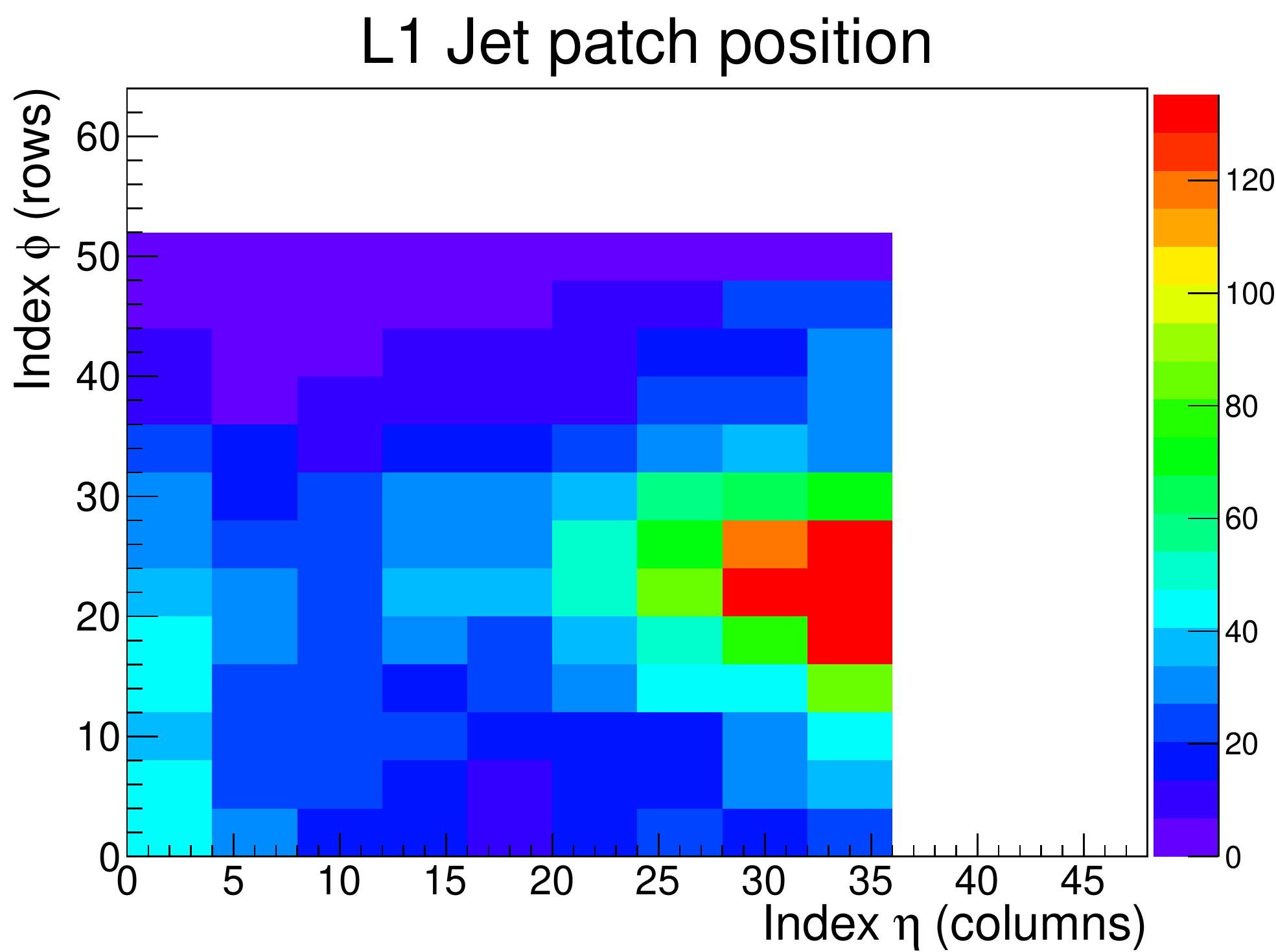}
  \includegraphics[angle=0,width=0.45\textwidth]{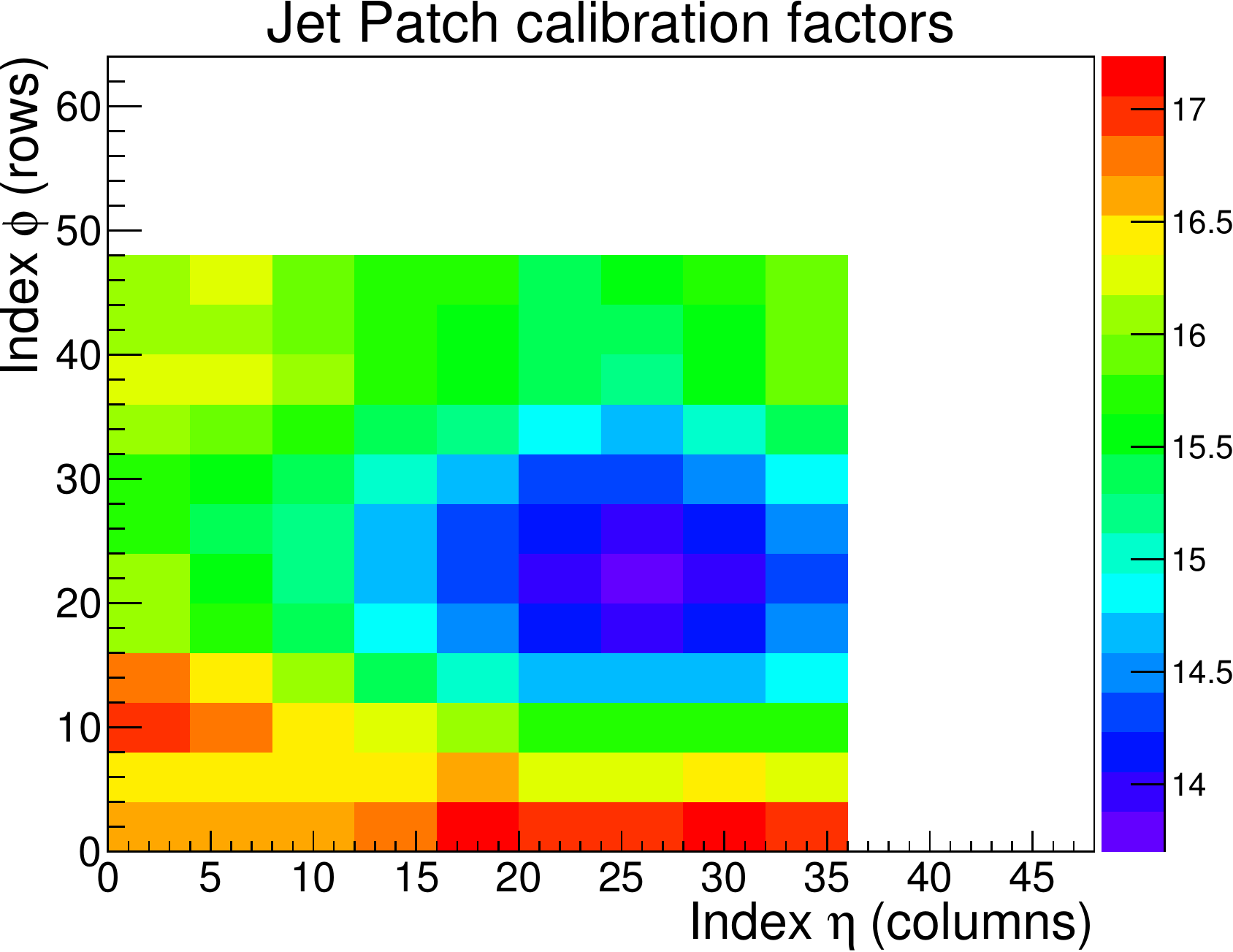}
  \caption{Left plot shows the occurrence of a jet trigger patch, a factor of six can be observed between the most active and the least active patch. Right plot shows in-situ calibration constants using $\pi^0$  for each jet trigger patch.}
  \label{spatial_uniformity}
  \end{center}
\end{figure}

As shown on the left of fig.~\ref{spatial_uniformity}, a spatial non uniformity in jet triggers was observed.
While the APD inter-calibration was done in the laboratory using cosmics, an in-situ calibration was performed using $\pi^0$ data at the end of 2011.
The calibration constants obtained roughly reproduce the trigger non uniformity.
This modified APD inter-calibration correction was used in 2012, further detailed analysis are required to assess the correction benefit.

\section{Perspectives}
Thanks to STU flexibility (available FPGA resources and spare trigger outputs), a 2\textsuperscript{nd} set of threshold parameters has been implemented to improve the L1 data sample overlap with the MB data sample. The resource usage increased from 69\% to 92\%.  
In mid-2013, ALICE is forecasted to be upgraded with the Di-jet CALorimeter, that will increase the coverage by $\Delta \eta \times \Delta \phi = 1.4 \times 100$\textdegree (PHOS included). 
It is composed of six shorter SM that will be installed on either side of the PHOS. For this operation one or two STUs will be used, depending whether PHOS will be included in the DCAL trigger or not (one STU for DCAL, one STU for PHOS). 

\section{Summary}
The STU has been installed for two years, and all the system interfaces have been validated.
The custom serial protocol, which has been modified, has been demonstrated to operate in realistic conditions with intensive readout. 
The fast FPGA remote configuration proved to be an asset for regular upgrades and problem solving. Moreover, it has been noted that from early on, it is an advantage to implement monitoring tools and to develop diagnosis tools. For instance, the ``event player'' demonstrated to be good tool for in situ validation of trigger algorithm without beam.


\begin{thebibliography}{9}

\bibitem{ALICEref} The ALICE Collaboration et al, The ALICE experiment at the CERN LHC, \href{http://dx.doi.org/10.1088/1748-0221/3/08/S08002}{2008 JINST 3 S08002}
\bibitem{LHCref} Lyndon Evans and Philip Bryant, LHC machine, \href{http://dx.doi.org/10.1088/1748-0221/3/08/S08001}{2008 JINST 3 S08001}
\bibitem{EMCALTDR} EMCal Technical design report, \href{http://cdsweb.cern.ch/record/1121574}{CERN-LHCC-2008-014}
\bibitem{FEEpaper} H. Muller et al., Front-end electronics for the ALICE calorimeters, \href{http://dx.doi.org/10.1016/j.nima.2009.09.022}{NIM A, vol. 617, Issues 1-3, p369}
\bibitem{TRUpaper} H. Muller et al., Hierarchical trigger of the ALICE calorimeters, \href{http://dx.doi.org/10.1016/j.nima.2009.06.097}{NIM A, vol. 617, Issues 1-3, p344}
\bibitem{V0paper} Technical Design Report on Forward Detectors FMD, T0 and V0, \href{https://edms.cern.ch/document/498253/1}{\emph{CERN-LHCC-2004-025}}
\bibitem{DDLpaper} DDL documentation web site \href{http://alice-proj-ddl.web.cern.ch/alice-proj-ddl/}{http://alice-proj-ddl.web.cern.ch/alice-proj-ddl/}
\bibitem{ALICEDAQ} ALICE trigger data-acquisition high-level trigger and control system, \href{http://cdsweb.cern.ch/record/684651}{\emph{CERN-LHCC-2003-062}}
\bibitem{CTPpaper} Central Trigger Process web site \href{http://epweb2.ph.bham.ac.uk/user/krivda/alice/}{http://epweb2.ph.bham.ac.uk/user/krivda/alice/}
\bibitem{TTCpaper} TTC documentation web site \href{http://ttc.web.cern.ch/TTC/intro.html}{http://ttc.web.cern.ch/TTC/intro.html}
\bibitem{STU_twepp2010} O. Bourrion et al, Level-1 jet trigger hardware for the ALICE electromagnetic calorimeter at LHC, \href{doi:10.1088/1748-0221/5/12/C12048}{2010 JINST 5 C12048}
\end{thebibliography}
\end{document}